\documentclass[sigconf,natbib=false, screen]{acmart}
\acmConference[ESEC/FSE 2023]{The 31st ACM Joint European Software Engineering Conference and Symposium on the Foundations of Software Engineering}{11 - 17 November, 2023}{San Francisco, USA}
\AtBeginDocument{%
  \providecommand\BibTeX{{%
    \normalfont B\kern-0.5em{\scshape i\kern-0.25em b}\kern-0.8em\TeX}}}
\setcopyright{acmcopyright}
\copyrightyear{2023}
\acmYear{2023}
\acmDOI{XXXXXXX.XXXXXXX}

\acmConference[ESEC/FSE 2023]{The 31st ACM Joint European Software Engineering Conference and Symposium on the Foundations of Software Engineering}{11 - 17 November, 2023}{San Francisco, USA}

\RequirePackage[
  datamodel=acmdatamodel,
  style=acmnumeric,
  ]{biblatex}

\usepackage{microtype}
\usepackage{hyperref}
\usepackage{subcaption}
\usepackage{multirow}
\usepackage{booktabs}
\usepackage{tcolorbox}

\begin{document}


\title{Evaluating Transfer Learning for Simplifying GitHub READMEs}
 \author{Haoyu Gao}
\affiliation{
 \institution{The University of Melbourne}
    \city{Melbourne}
  \state{Victoria}
  \country{Australia}
 }
 \email{haoyug1@student.unimelb.edu.au}

\author{Christoph Treude}
\affiliation{
\institution{The University of Melbourne}
    \city{Melbourne}
  \state{Victoria}
  \country{Australia}
}
\email{christoph.treude@unimelb.edu.au}

\author{Mansooreh Zahedi}
\affiliation{
\institution{The University of Melbourne}
    \city{Melbourne}
  \state{Victoria}
  \country{Australia}
}
\email{mansooreh.zahedi@unimelb.edu.au}

\begin{abstract}
Software documentation captures detailed knowledge about a software product, e.g., code, technologies, and design. It plays an important role in the coordination of development teams and in conveying ideas to various stakeholders. However, software documentation can be hard to comprehend if it is written with jargon and complicated sentence structure. In this study, we explored the potential of text simplification techniques in the domain of software engineering to automatically simplify GitHub README files. 
We collected software-related pairs of GitHub README files consisting of 14,588 entries, aligned difficult sentences with their simplified counterparts, and trained a Transformer-based model to automatically simplify difficult versions. To mitigate the sparse and noisy nature of the software-related simplification dataset, we applied general text simplification knowledge to this field. Since many general-domain difficult-to-simple Wikipedia document pairs are already publicly available, we explored the potential of transfer learning by first training the model on the Wikipedia data and then fine-tuning it on the README data. Using automated BLEU scores and human evaluation, we compared the performance of different transfer learning schemes and the baseline models without transfer learning. The transfer learning model using the best checkpoint trained on a general topic corpus achieved the best performance of 34.68 BLEU score and statistically significantly higher human annotation scores compared to the rest of the schemes and baselines. We conclude that using transfer learning is a promising direction to circumvent the lack of data and drift style problem in software README files simplification and achieved a better trade-off between simplification and preservation of meaning.
\end{abstract}



\maketitle

\section{Introduction}

Software documents describe key information about software products, such as technologies, code structure, system design, and architecture. These documents are an integral part of the software development process~\cite{parnas2009document} as they can be used to describe the application requirements, design decisions, architecture, as well as deployment and installation. 
In particular, README files shape developers' first impression about a software repository and document the software project that the repository hosts~\cite{prana2019categorizing}. 
However, README documents often contain jargon, abbreviations, and code blocks, making the text challenging to comprehend for non-specialists and people from other language backgrounds. In fact, readability issues and complicated documents are important issues that practitioners frequently encounter~\cite{aghajani2020software}. A simple search on GitHub for ``complicated README'' yields over 27,000 issues and 37,000 pull requests\footnote{\url{https://github.com/search?q=complicated+readme&type=issues}}. Therefore, simplification of README files is needed to improve the efficiency of communication between members of development teams and to propagate new technologies to broader fields. 

Significant advancement in Natural Language Processing (NLP) has been witnessed over the last decade, thanks to the development of artificial neural networks. Text Simplification (TS) is an NLP task that focuses on rewriting texts into simpler versions while preserving the original meaning to the extent possible. The simplification of text in the general domain has been studied extensively, and its data sources include mainly Wikipedia and its ``Simple-Wikipedia'' counterpart~\cite{napoles2010learning} as well as Newsela articles written for specific age groups~\cite{xu2015problems}. These data sources, especially their simplified versions, are written by professionals with the intention of catering to people with different levels of reading ability. The simplification of sentences in general domain text can be implemented by three main operations, including splitting, deletion, and paraphrasing~\cite{feng2008text}. These simplification operations are implicitly encoded in text simplification datasets of the general domain, and neural simplification models trained on these datasets memorise the rules in their parameters and achieve competitive performance~\cite{zhao2018integrating, feng2017memory}.

The significant disparity between the simplification of general domain documents and domain-specific software documentation prohibits the simple reuse of text simplification systems designed for the general domain. First, the text style of software documentation includes many code blocks and external links such as URLs, which do not resemble the style for general-domain texts. In addition, the simplification rules for software documents differ from the general-domain text by performing fewer deletions and more elaborations. 
Indeed, in this paper we show that a state-of-the-art approach trained on general-domain documents from Wikipedia is not able to produce semantically identical and/or grammatically correct content in around 80\% of the cases when applied to README files (cf.~Section \ref{human anno}).


A simplification system trained directly on general-domain text cannot simplify software documentation satisfactorily. To the best of our knowledge, there is no previous research focusing on simplifying software documentation. To address this gap, we collected a software-specific text simplification dataset and trained a simplification system on the data. We experimented with transferring general-domain text simplification rules to software documentation and evaluated the system through automatic metrics and manual analysis. We found that by applying transfer learning, the model was able to generate the most satisfactory simplifications. Specifically, the best-performing model achieved a 34.68 BLEU score in the test set and exceeded the baseline models in terms of semantic similarity, grammar, and simplicity, based on human annotation.

The contributions of our research include the following: 
\begin{itemize}
    \item A README files simplification dataset.
    \item A pipeline to collect such a dataset.
    \item An exploration of the application of transfer learning to the problem of simplification of README files with promising results.
\end{itemize}


%

\section{Related Work}
\subsection{Documentation Issues and Solutions}
Previous studies provide rich empirical evidence for software documentation issues. Steinmacher et al.~\cite{steinmacher2015social} discovered several barriers to participating in Open Source Software (OSS) projects, one related to ``Documentation problems''. After that, Aghajani et al.~\cite{aghajani2019software, aghajani2020software} conducted empirical studies to investigate software documentation issues and practitioners' perspectives. Software documentation issues could generally be categorised into what information is contained and how the information is presented.

To address the problems, the automatic generation of software documentation could potentially mitigate correctness, completeness, up-to-dateness and various other issues. Automatic software documentation generation can be applied to various software artifacts, including source code~\cite{sridhara2010towards, moreno2013automatic, hu2018deep}, bug reports~\cite{rastkar2010summarizing, rastkar2014automatic} and pull requests~\cite{liu2019automatic}. In terms of source code, Sridhara et al.~\cite{sridhara2010towards} used algorithms to generate comments for Java methods, while McBurney and McMillan~\cite{mcburney2014automatic} improved it by adding surrounding contexts. Moreno et al.~\cite{moreno2013automatic} summarised Java classes using stereotype rules and manually defined templates. Hu et al.~\cite{hu2018deep} proposed using a sequence-to-sequence model and formulated code summarisation as a translation task.

In terms of bug reports, Rastkar et al.~\cite{rastkar2010summarizing, rastkar2014automatic} trained a classifier to identify important sentences from bug reports and used them as summaries. Regarding API documentation, Treude and Robillard~\cite{treude2016augmenting} augmented API documents using insight sentences from Stack Overflow. Pull request descriptions can also be generated considering commits and code comments~\cite{liu2019automatic}. Source code changes are also used to generate commit messages~\cite{xu2019commit, dong2022fira}.

Automatic documentation generation could help developers identify components that are prone to be overlooked and improve development efficiency. Among the issues, readability is an important issue that practitioners frequently encounter~\cite{aghajani2020software}. One of the practitioners in Aghajani et al.'s survey stated, ``A developer in our team created confusing and overly complicated documentation for customers of our solution''. To the best of our knowledge, no previous studies focused on simplifying software documentation to improve people's comprehension. Text simplification is an NLP technique that could bridge this gap and enhance developers' understanding of software documentation.

\vspace{-0.4cm}
\subsection{Text Simplification}
Multiple data sources have been proposed for the task of simplification of text. Zhu et al.~\cite{zhu2010monolingual} first used Wikipedia and Simple Wikipedia as a source, which was later expanded by Zhang and Lapta~\cite{zhang2017sentence} to the WikiLarge dataset. However, Xu et al.~\cite{xu2015problems} argued that this Wikipedia-based simplification dataset is suboptimal and difficult to generalise to other genres of text. They proposed a new dataset called Newsela, which contains different levels of simplification. Moreover, there are also simplification corpora of languages other than English 
\cite{brunato-etal-2015-design, gala-etal-2020-alector, specia2010translating, vstajner2015automatic}. Due to easier access to a large corpus of Wikipedia data, we performed part of our experiment on Wikipedia datasets.

Meanwhile, the success of a text simplification system is highly dependent on the quality and quantity of complex-simple sentence pairs in the training corpus~\cite{jiang2020neural}. Zhu et al.~\cite{zhu2010monolingual} first used sentence-level TF-IDF (term frequency inverse document frequency) similarity to construct the alignment between a simple Wikipedia corpus and its regular counterpart. Later, more sophisticated sentence alignment techniques were proposed that consider sentence orders and word-level similarity
\cite{coster-kauchak-2011-learning, woodsend-lapata-2011-learning, hwang2015aligning, kajiwara-komachi-2016-building}, increasing the alignment quality and the dataset size.
Recently, Jiang et al.~\cite{jiang2020neural} proposed a neural-based Conditional Random Field (CRF) aligner, which decomposes the potential function into semantic similarity (approximated by the BERT classifier) and alignment label transition (approximated by the feedforward network). Their model automatically aligns 604k non-identical aligned and partially aligned sentence pairs. This powerful tool is able to achieve more than 0.9 F1 score on the previous Wikipedia corpus alignment task, thus making their auto-aligned dataset of higher quality. Sentence alignment is the first procedure in the pipeline of text simplification. In our research, we borrow their idea to align software document pairs as our first step in building a software documentation simplification system.

Recent work began to see text simplification as a monolingual translation task. Specia~\cite{specia2010translating} first applied statistical machine translation to text simplification. Kauchak~\cite{kauchak2013improving} incorporated regular and simple sentences to train an n-gram language model to perform text simplification tasks. Nisioi et al.~\cite{nisioi2017exploring} began to see text simplification as a task similar to machine translation (MT) and trained a standard sequence-to-sequence model based on LSTM that surpasses the performance of previous statistical MT models. Different network designs were also developed for the model to learn a more effective simplification. Zhang and Lapta~\cite{zhang2017sentence} used reinforcement learning for simplification with a reward that approximates simplicity, relevance, and fluency. Zhang et al.~\cite{zhang2017constrained} combine lexical simplification with sentence-level simplification by first performing lexical substitution and then feeding the sentence into a constrained sequence-to-sequence model. Nishihara et al.~\cite{nishihara2019controllable} proposed a controllable simplification system by adding a level token and modifying the loss function, while Mallinson and Lapata~\cite{mallinson2019controllable} did it employing syntactic and lexical constraints.  
Current text simplification systems use transformer architecture~\cite{jiang2020neural} and can achieve state-of-the-art performance. 
In our work, we primarily used the transformer model and explored
the simplification rule gap between software and general-domain documentation.

When there is a disparity between the data distribution (such as the text styles for software documentation and documents of the general domain), the performance of the model can be degraded~\cite{shimodaira2000improving}, in which case transfer learning is needed. Transfer learning improves the performance of a learner in one domain by transferring information from a related domain~\cite{weiss2016survey}. It is widely used in many areas, including image processing~\cite{resnet} and natural language processing~\cite{devlin2018bert}, and has achieved significant success. In our work, we experimented with various transfer learning techniques for the task of software documentation simplification. We applied the knowledge learnt from general-domain document simplification to mitigate the noisy and sparse attributes of software-related texts.



\section{Data Collection}
In order to obtain enough software-related documents to train our model, we collected README files from GitHub using its RESTful API. We implemented a program using GitHub access tokens to iterate from the very first repository ordered by GitHub id, and check for candidates for the simplification dataset. We only considered repositories not forking other repositories and with at least ten stars 
to filter out toy projects. One reason we collected older repositories is that we believe README files in older repositories need more simplification, as different techniques were used back then, and more old repositories have gone through simplification updates compared to recent repositories. As we needed to get updates in the READMEs, longer commit histories will be more likely to contain candidate data. Therefore, only projects with at least 100 commits are investigated.  The left part of Figure~\ref{fig:procedure} describes the procedures for collecting the data.

Specifically, for each repository, we iterated through its entire commit history. We collected a list of keywords that can be a hint for simplification. We identified those commits that contain at least one of those keywords and only modify the README file as document simplification instances. The previous README file is marked as the difficult version, and the newly committed file is marked as the simplified version. To encourage more prominent simplifications and avoid training data duplication, we only preserved the first commit and the last commit with simplifications on the README file for each repository. We collected 14,588 document-level regular-to-simple instances in total.

Regarding keyword selection, we initially chose three keywords, i.e., ``simplify'', ``clarify'', and ``explain''. Their definitions were searched in WordNet~\cite{miller1990introduction}, and their synonyms were further added to the set of keywords. After that, we expanded the keyword set by adding different forms, including nouns, verbs, and adjectives. The keywords, along with their distribution in the final collected corpus, are listed in Table~\ref{tab:keywords}. Looking at the table, the keywords ``clarify'', ``simplify'', ``explain'' and ``ease'' along with their derivations are the most frequently used keywords in the collected documents. The more complicated words like ``elucidate'' and ``comprehend'' were rarely used. To provide readers with more information on the effect of these keywords, we also added sample commit messages with the most common word sets and listed them in Table~\ref{tab:keywords}.

\begin{table}[t]
    \centering
    \caption{List of Keywords and their Distribution in Data}
    \begin{tabular}{lrrl}
    \toprule
    keywords & count & sum & sample commit message\\
    \midrule
    simplification & 51  & \multirow{4}{*}{2,756} & \multirow{4}{*}{\shortstack[l]{Simplify intro paragraph}}\\
    simplify & 1,524 & \\
    simple & 1,161 & \\
    simplicity & 20 & \\
    \midrule
    reduction & 20 & \multirow{2}{*}{314} & \multirow{2}{*}{\shortstack[l]{Change link text to\\ reduce confusion}} \\
    reduce & 294 & \\
    \midrule
    clarification & 954 & \multirow{4}{*}{7,039} & \multirow{4}{*}{Clarifying README a bit}\\
    clarify & 3,924 & \\ 
    clear & 1,677 & \\
    clarity & 484 & \\
    \midrule
    elucidation & 1 & \multirow{4}{*}{2} & \multirow{4}{*}{\shortstack[l]{Elucidate what we do\\ with errorCode.}}\\
    elucidate & 1 & \\
    elucidative & 0 & \\
    elucidatory & 0 & \\
    \midrule
    explanation & 1,412 & \multirow{3}{*}{3,419} & \multirow{3}{*}{\shortstack[l]{Update the documentation\\ to explain how this works}}\\
    explain & 1,983 & \\
    explanatory & 24 & \\
    \midrule
    comprehension & 10 & \multirow{3}{*}{14} & \multirow{3}{*}{more comprehensible} \\
    comprehend & 1 & \\
    comprehensible & 3 & \\ 
    \midrule
    ease & 46 & \multirow{2}{*}{1,044} & \multirow{2}{*}{\shortstack[l]{Rewrote README.md to make\\it easier to follow}}\\
    easy & 998 & \\
    \bottomrule
    \end{tabular}
    \label{tab:keywords}
\end{table}

Although keywords in commit message histories convey information about the modified contents, using a unigram of occurrence can be ambiguous. For example, a simple negation term could render the semantics of simplification to the opposite meaning. Also, sometimes the hint word for simplification might not refer to the harvested README file, but to structures in the code blocks. Furthermore, even if it refers to the simplification of the README file, only a few sentences might be simplified, with most parts remaining unchanged. Therefore, further filtering and preprocessing steps are required, which are illustrated in Section 4.

In terms of implementation detail, we used the authors' access tokens, and use PyDriller~\cite{spadini2018pydriller} to mitigate the impact of the GitHub RESTful API rate limit as much as possible. 
The collected documents are in JSON format, with fields including difficult document, simple document, commit message, language used, and project fork counts. We collected 14,588 of these document pairs in total, which are used to construct our software document simplification dataset.

\begin{figure}[h!]
    \centering
\includegraphics[scale=0.45]{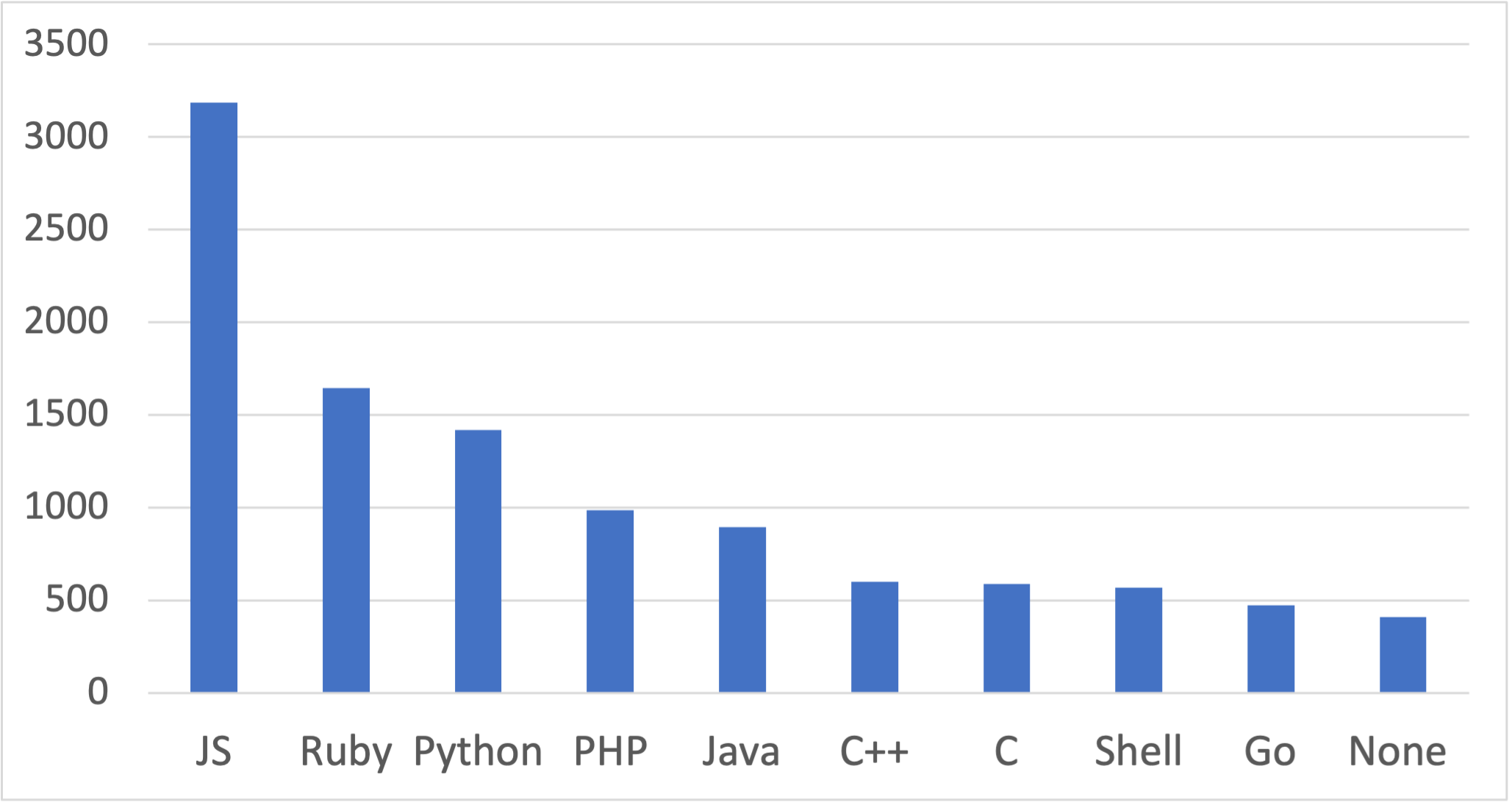}
    \caption{Repository Languages}
    \label{fig:repo_lang}
\end{figure}

Instead of focusing on certain programming languages, we want to investigate the overall simplification of README files through the commit history and thus did not filter on the programming language field. Figure \ref{fig:repo_lang} lists the top ten languages used by the repositories that we collected.

\section{Data Preprocessing}
\begin{figure}[h!]
    \centering
    \includegraphics[scale=0.35]{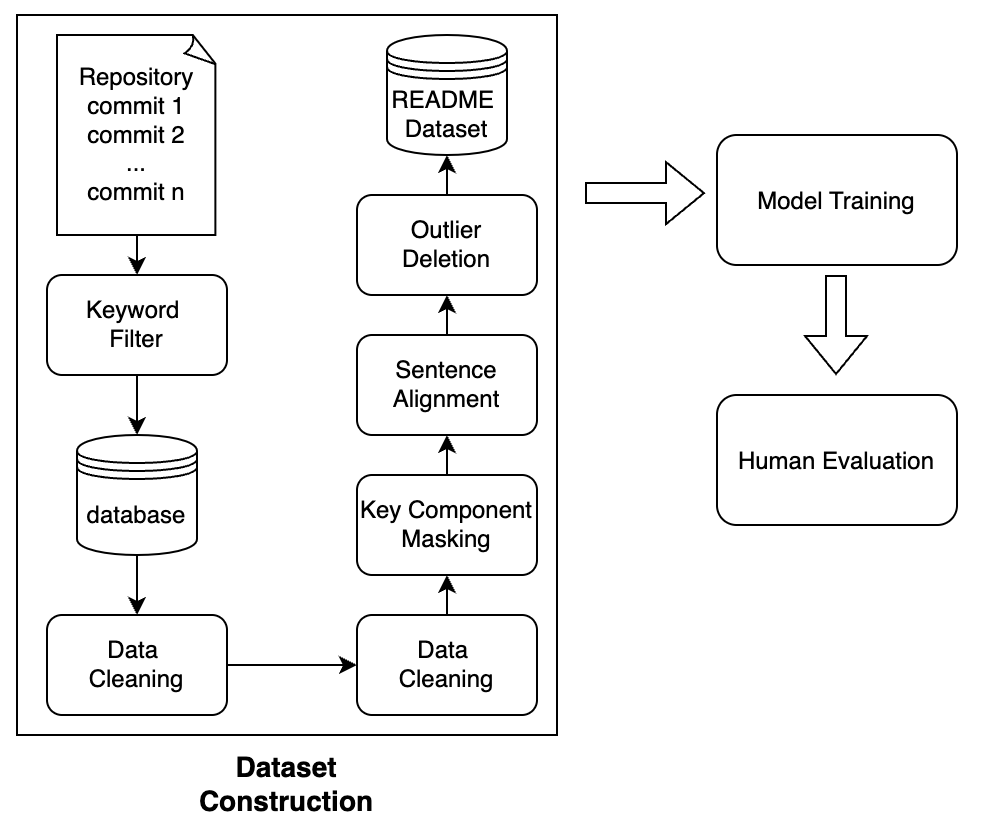}
    \caption{Overall Procedure}
    \label{fig:procedure}
\end{figure}

After harvesting  GitHub README files, the difficult-to-simple correspondence is at the document level. This is too long to build an effective translation model directly, as most sentences in two versions of documents are duplicates, which creates difficulty for the model to learn simplification. Also, considering that it is not reliable to only depend on heuristic keywords as an indicative sign of simplification, we need to further filter the harvested dataset and perform the sentence alignment task in order to give a higher confidence dataset in a sentence-level correspondence. Figure~\ref{fig:procedure} depicts the overall procedure, with the left hand side describing the dataset construction process. Each component will be discussed in detail in this section.

\subsection{Data Cleansing and Masking}

\begin{table}[h!]
    \caption{Semantic Components and assigned Tokens}
    \centering
    \begin{tabular}{ll}
    \toprule
    \textbf{Component Type} & \textbf{Token} \\
    \midrule
    inline code block  & $ \langle code\_small \rangle $ \\
    chunks of code & $ \langle code\_large \rangle $ \\ 
    path of file or directory & $ \langle file \rangle $\\
    table &  $ \langle table \rangle $ \\
    hyperlink & $ \langle url \rangle $ \\
    \bottomrule
    \end{tabular}
    \label{tab:special_tokens}
\end{table}

The collected README files are written in several formats, including recommended markdown style\footnote{\url{https://docs.github.com/en/get-started/writing-on-github/getting-started-with-writing-and-formatting-on-github/quickstart-for-writing-on-github}}, plain text and HTML syntax, which makes them noisy. For example, some people follow the conventions ``-'' and ``|'' for constructing tables, while others choose to use HTML syntax. For data cleaning and preprocessing, we first removed emojis and different spacing characters including ``\textbackslash t'' and ``\textbackslash n''.

Another critical characteristic of our data source is that it contains a large number of semantic components that depend on the document and its context. These components include URLs, code blocks, tables, etc. These components are essential for software documentation, as they usually include instructions, specifications, and external links that elaborate on projects. But it can be challenging for the translation model to implicitly learn their attributes as different documents typically contain components that are barely the same. Therefore, we used Python package ``markdown2''\footnote{\url{https://github.com/trentm/python-markdown2}} to identify and convert these special components into different individual tokens that are distinguishable for their usage. The translation model is explicitly told where the special components are and can generate more cohesive sentences. We manually inspect the data and categorise the tokens into five types. The special tokens are listed in Table~\ref{tab:special_tokens}. 

Version requirements and plain text code without using the markdown syntax are also important semantic information in the sentences. However, detecting these elements would require using regular expressions, which is noisy and not the major goal of this paper. Therefore, we leave these elements as in their original form.

\subsection{Sentence Alignment}
Sentence alignment methods were extensively studied in previous research. Jiang et al.~\cite{jiang2020neural} recently developed a neural-based CRF sentence alignment method that achieves an F1 score over 0.9 on Wikipedia data. They decompose the potential function as follows:
$\psi(a_{i}, a_{i-1}, S, C)=sim(s_{i}, c_{a_{i}})+T(a_{i}, a_{i-1})$, 
where S denotes simple sentences, C denotes complex sentences and $a_{i}$ denotes the index of the aligned sentence. A fine-tuned BERT model is used to approximate $sim(s_{i}, c_{a_{i}})$, and a simple multi-layer perceptron is used to approximate $T(a_{i}, a_{i-1})$. They finally used a Viterbi algorithm for decoding the optimal alignment arrangements.

For Wikipedia data, since original and simple documents are not composed concurrently, the positions of difficult sentences and their simplified correspondence could differ a lot. However, the aligned sentences tend to be in a relatively similar order in terms of our harvested GitHub README files due to the incremental development nature of many software projects. Therefore, the calculation of $T(a_{i}, a_{i-1})$ in our software documents will not benefit much and will only increase training and decoding time. Therefore, we discarded other components and only borrowed the fine-tuned BERT classifier to perform our alignment task.

The specific alignment task is performed as follows. For each pair of regular-simple documents, the sentences of the simplified document will be fed into the BERT classifier with the regular sentences one by one. For those that are classified as ``aligned'', we would temporarily mark them as aligning candidates. To avoid $O(n^2)$ time complexity when performing the alignment task, we exploited the fact that many GitHub README files tend to grow incrementally. Unless a complete refactor of the documents, aligning sentences should appear at the a closer section compared to non-aligning ones. Each sentence in the simplified document will only be compared with the regular ones that have the sentence position within a window size of 50 to the simplified sentence. This window size is able to cover most of the README file sentences, except for excessively long ones, thus reducing the processing time while providing good coverage for the majority of sentence pairs.  However, the drifted sentence style for software documentation and the large amount of potential matches for each simple sentence make the false positive rate relatively high. To accommodate this situation, multiple filtering methods rules are used.

First, we filtered the candidates using the TF-IDF-based cosine distance. TF-IDF is a commonly used statistic in natural language processing, which computes weights for each occurring word by taking into account the frequency of the word as well as the frequency of documents containing it. In this case, it does not give great weight to frequently occurring but meaningless words such as ``the'' and ``a''. We trained our TF-IDF model on the corpus of all README files.  Using the TF-IDF vectorizer, each sentence is represented as a vector, and we are able to compute the cosine distances between simple and complicated sentences.

\begin{figure}[h!]
    \centering
    \includegraphics[scale=0.3]{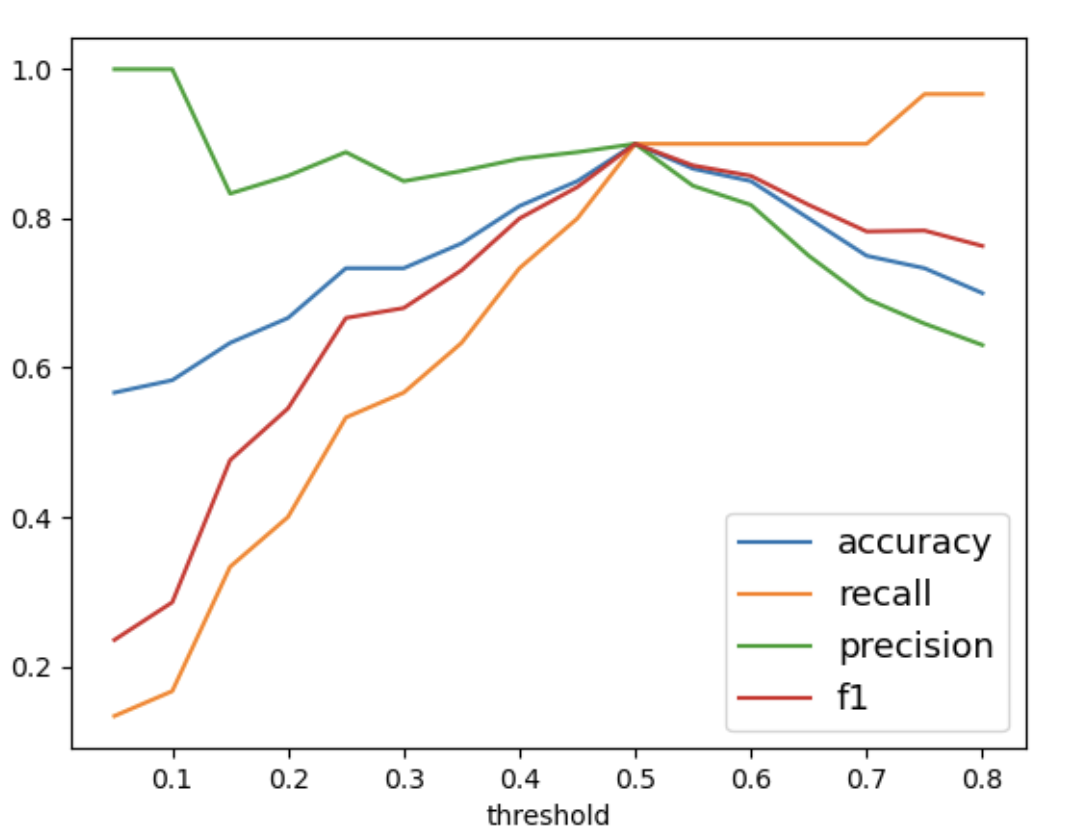}
    \caption{Performance of Different Thresholds}
    \label{fig:cosine distnce}
\end{figure}

To filter out false positive candidate pairs using the cosine distance based on TF-IDF, we manually selected 60 simple to complicated sentence pairs and labelled them with the ground truth alignments, with 30 pairs labelled ``aligned'' and the other 30 as ``not aligned''. We experimented with different filtering thresholds for cosine distances, and the result is shown in Figure~\ref{fig:cosine distnce}.

As seen in the figure, the F1 score and recall increase until the cosine distance threshold of 0.5. After that, the recall flattens while both the precision and F1 score start to decrease. As in the real collected alignment candidate pairs, more of them tend to be false positive instead of only taking half the proportion, we choose the threshold of 0.5 to prevent a further drop in the accuracy score. All candidate pairs with TF-IDF cosine distances greater than 0.5 are categorised as false positives and discarded in this step.

Furthermore, the BLEU score~\cite{papineni2002bleu} is a widely used metric for machine translation tasks that computes the n-gram overlaps between the target and reference sentences and provides an intuitive score for the level of similarity between sentences. A BLEU score greater than 0.9 typically indicates merely a copy of the source sentence, while a BLEU score below 0.1 means overlap only in some name entities~\cite{jiang2020neural}. Therefore, we discarded sentence pairs with BLEU scores greater than 0.9 or less than 0.1. The size of the dataset after applying the TF-IDF distance 
and BLEU score filter is 43,772. 

\subsection{Dataset Anomaly Filter}
After performing the previous alignment steps, we have constructed a dataset of regular-to-simple software documents. To obtain a high-quality dataset, we collected statistics on the number of alignments of regular sentences for each simple sentence and eliminated those that appeared to be outliers. The average alignment number for the simplified sentence is 1.2, with a maximum number of 40 and a variance of 0.8. We then removed data that are outside the range of $3\sigma$. As a result, only sentences that are aligned with no more than three sentences were preserved. We also discarded excessively long sentences. Sentences with more than 40 alphabetic words were eliminated. This procedure further reduces the size of the dataset to 34,667.




After a closer look at these eliminated sentence pairs that were initially categorised as ``aligned'' by our sentence aligner, most of these outliers either appear to be too short or repeat instructions that only change a few words or URLs. For example, an original document of ``$[video](\langle url \rangle)](\langle url\rangle)$'', which is a markdown syntax to show some URLs, is matched with three other texts of ``$[\langle url\rangle](\langle url\rangle)](\langle url\rangle)$'' in the simplified document. These masked sentences are short but similar only in markdown syntax structure instead of semantic meanings, and should be considered as noise. Using this method, we further cleaned up our proposed dataset.
\subsection{Dataset Comparison}
For the simplicity of elaboration, we refer to the dataset constructed using the Wikipedia and Simple Wikipedia source as ``wiki-data'' and refer to the dataset we constructed in the previous steps as ``sw-data''. In this section, we briefly discuss the attributes of both datasets. Table~\ref{tab:data statistics} lists statistics for ``wiki-data'' and ``sw-data''.

\begin{table}[h]
    \caption{Statistics for sw-data}
    \centering
        \begin{tabular}{lrrr}
            \toprule
               & Simple & Regular & Simple-Regular Ratio \\
            \midrule
            \multicolumn{4}{l}{sw-data statistics}\\
            \midrule
            Average Length & 24.80 & 26.62 & 93.2\% \\
            Vocabulary Size & 21,653 & 22,313 & 97.0\% \\
            Exclusive Vocab Size & 2,889 & 3,549 & 81.4\% \\
            \midrule
            \multicolumn{4}{l}{wiki-data statistics}\\
            \midrule
            Average Length & 14.76 & 20.91 & 70.6\%  \\
            Vocabulary Size & 32,228 & 37,278 & 86.5\% \\
            Exclusive Vocab Size & 1,171 & 6,221 & 18.8\%\\
            \bottomrule
        \end{tabular}
        \label{tab:data statistics}
\end{table}

As seen from the table, these two datasets differ significantly. Specifically, sentences in wiki-data tend to have a shorter length. An average length of over 24.80 and 26.62 for simple and regular sentences in the sw-data indicates that sentences harvested from GitHub can be more complex and wordy. Moreover, the vocabulary size in wiki-data significantly surpasses that of the sw-data. This can happen because software documentation only focuses on specific topics, while wiki-data covers a much wider range of topics.

Also, the simple-to-regular ratio statistics in wiki-data and sw-data indicate that the simplification in wiki-data is more aggressive. In contrast, the simplification of sw-data makes less apparent changes. This could happen because Simple Wikipedia articles are written with the intention of letting non-native speakers feel confident in reading. At the same time, the simplification in GitHub files includes different operations such as rewrite, exemplify and clarify, and some of the detail changes are minor. As the sw-data dataset contains relatively less apparent simplification, the model tends to memorise the original sentence and barely performs simplification. The simplification rule gap between wiki-data and sw-data, plus the noise in the sw-data, motivates us to explore transfer learning, as discussed in the next section.

To further illustrate our points, we picked two representative simplification examples, one from wiki-data and the other from sw-data, as can be seen in Table~\ref{tab:simplification examples}. In this example, the wiki-data simplification performs aggressively and ignores some details of the evolution of the presidency armies. However, the author who simplified the sw-data document only changes a few words at the end of the sentence, making the argument clearer by giving a specific instruction.

\begin{table*}[t]
    \caption{Simplification Examples}
    \centering
    \begin{tabular}{lp{8cm}p{7.7cm}}
    \toprule
         & regular & simple \\
         \midrule
        sw-data &   \#\# limitations * due to the nature of irssi's readline, it is not possible to add formatting directly in the input line, hence the need for the extra window kludge.  &  \#\# limitations * due to the nature of irssi's readline, it is not possible to add formatting directly in the input line, so an extra line is output to the screen instead.  \\
        \midrule
        wiki-data & The presidency armies were the armies of the three presidencies of the East India Company's rule in India, later the forces of the British Crown in India, composed primarily of Indian sepoys.  &
        The presidency armies were the armies of the three presidencies of British India . \\
    \bottomrule
    \end{tabular}
    \label{tab:simplification examples}
\end{table*}

As we are going to use both datasets to train our software documentation simplification system, we split both datasets into train, validation, and test sets. For the sw-data, we have a train set of size 28,000, a validation set of size 3,500 and the rest of the data forms the test set. For the wiki-data, we have a train set of size 450,000 as well as a validation set and a test set both of size 20,000. The training of the model and the transfer learning will be conducted on the train set, and the performance of cross-entropy loss will be evaluated on the validation set. We will finally generate new texts on the test set for more detailed evaluation.

\section{Model Training and Transfer Learning}

In this section, we elaborate on how we
trained our model using transfer learning, and discuss the output of the model based on BLEU score evaluation.


\subsection{Model Tokeniser}
Before feeding sentences into our model, we need to tokenise sentences into a list of tokens so that the model can learn their representations in the embedding layer. The tokens can be whole words or subwords. A tokeniser off-the-shelf is able to perform well on general-domain tasks like simplifying wiki-data. However, software documentation has a lower lexical complexity and contains components that the model does not want to reduce. To better fit our study, a custom tokeniser is needed. Therefore, we trained our own tokeniser using all sentences in the sw-data and wiki-data training set using the WordPiece tokenisation algorithm~\cite{wu2016google}. WordPiece is a subword tokenisation algorithm developed by Google which is widely used in various models~\cite{devlin2018bert, sanh2019distilbert}. Similar to Byte Pair Encoding (BPE)~\cite{gage1994new}, it learns how to merge characters into words when provided with a large corpus. During tokenising, a sub-word with  a ``\#\#'' symbol indicates it is the continuation of the previous subword and is later concatenated.

In Section 4.1, we used regular expressions to mask these special components to prevent key components from automatic simplification to different tokens. These special tokens are listed in Table~\ref{tab:special_tokens}. However, to ensure that our tokeniser does not further split these tokens, we specify those as special tokens during our training of the tokeniser, along with $\langle sos \rangle$, $\langle eos \rangle$ and $\langle UNK \rangle$, indicating the start of a sentence, the end of a sentence and unknown words, respectively. In this case, the tokeniser can directly tokenise these components as a whole. As a result, the downstream model will directly know the meaning of these tokens, making it easier for the model to learn how to manipulate them. We also specified the vocabulary size of the tokeniser at 40,000.


\subsection{Model Architecture and Hyperparamters}
The text simplification task can be considered a translation task, in which sequence-to-sequence models are widely used. Transformer~\cite{transformer} is a multi-headed self-attention sequence-to-sequence model that achieves competitive performance in neural translation tasks. This architecture has become an essential building block in many models in the deep learning area. As this work focuses more on the simplification rules of software documentation and mitigating the drawbacks in the currently collected sw-data, we adopted the vanilla version of Transformer in the original paper~\cite{transformer} with only some minor changes in the tokeniser, and a reduction of trainable parameters to save training time. With limited access to GPU computing resources, plus the long training time of our model, we did not try to tune the hyperparameters extensively to reach the best performance. Instead, we experimented with only a few sets of hyperparameters close to the setting from ~\cite{jiang2020neural} and used one set of them that performs the best on the task of sw-data. This set of hyperparameters was later used on every model we trained, including the wiki-data and the transfer learning. Specifically, the model configuration and hyperparameter choices are listed in Table~\ref{tab:model configuration}.

\begin{table}[h!]
    \caption{Model Configuration and Hyperparameters}
    \centering
    \begin{tabular}{lr || lr}
    \toprule
        Multi-head numbers & 8 & Learning rate & 1e-5\\
        Encoder layers & 6 & Batch size & 8\\
        Decoder layers & 6 & Optimiser & Adam\\
        Embedding dimension & 512 & Alpha & 2e-5\\
        Feed-forward dimension & 2,048 & Dropout rate & 0.1 \\
    \bottomrule
    \end{tabular}
    \label{tab:model configuration}
\end{table}
\subsection{Training on wiki-data and sw-data}
\begin{figure}[h!]
    \centering
    \includegraphics[scale=0.25]{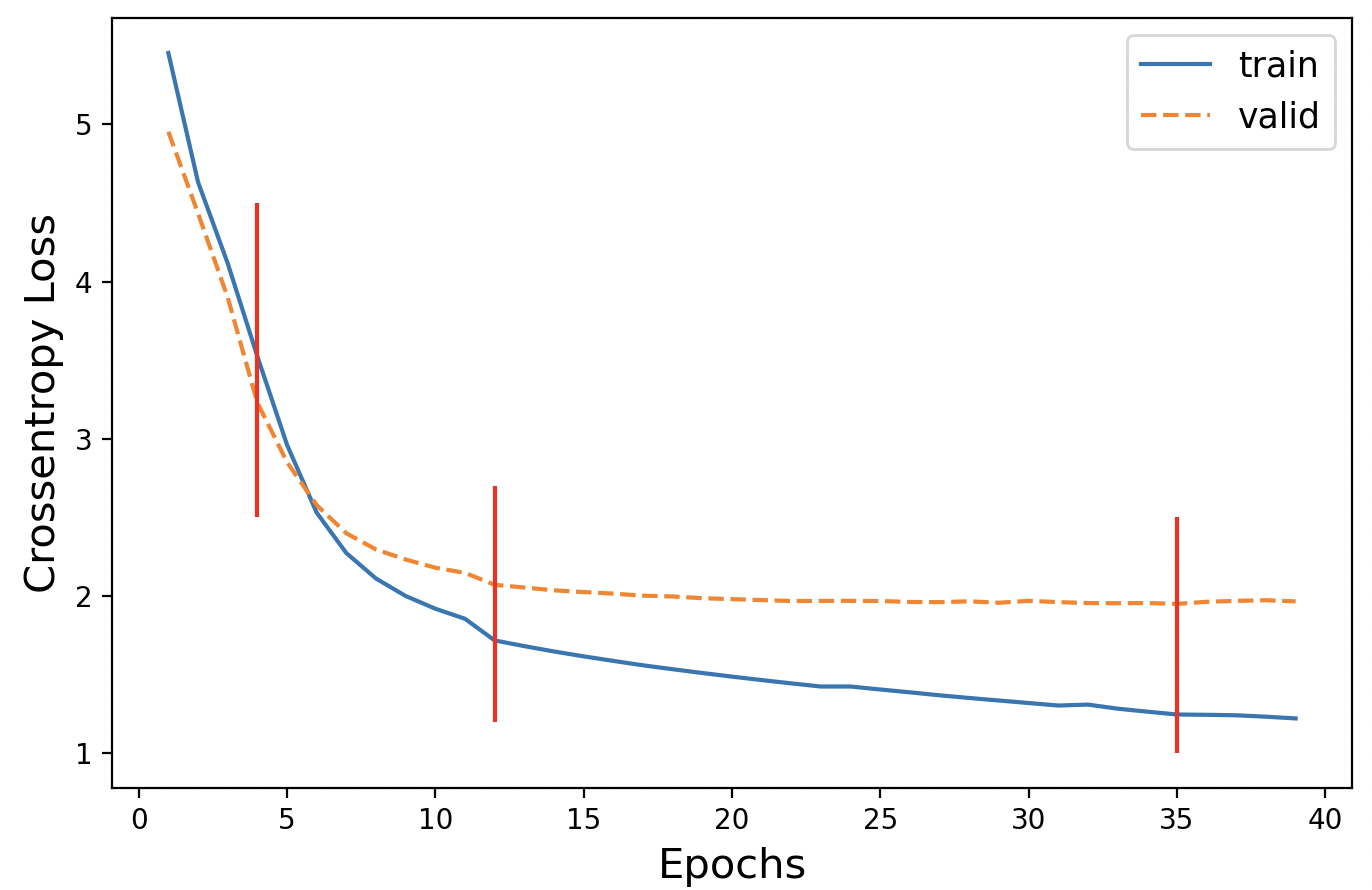}
    \caption{Wiki-data loss curve}
    \label{fig: wiki plot}
\end{figure}

As a starting point, models with the given architecture were trained solely on the wiki-data sets and the sw-data set. The cross-entropy loss curve of the entropy of the model trained with wiki-data is shown in Figure~\ref{fig: wiki plot}. and Figure~\ref{fig:transfer curves} 
 respectively.

We trained our models on wiki-data for 40 epochs and sw-data for 50 epochs. Their learning curves exhibit similar patterns representing the model gradually overfitted on the training set. In the final epochs, as the loss on the validation set is not decreasing, we stop the training process and preserve  checkpoints with the lowest validation error as final models. 

\subsection{Transfer Learning}
Because of the limitations mentioned in the Dataset Comparison section, it is difficult for the wiki-data-trained model to adapt to the change in text style and in simplification rule. Meanwhile, it is difficult for the sw-data-trained model to generate ideal simplifications, as the dataset contains different styles of simplification and many simplifications with only a few swaps of words. These two models are used as baselines to compare our later transferred learning models. For simplicity of argument, we denoted these two models as baseline wiki and baseline sw. A combination of both styles in wiki-data and sw-data, namely practical simplification and technical precision, is desired for software documentation simplification.  By applying transfer learning, we intend to share general-domain text simplification knowledge with the software documentation simplification task.

Figure~\ref{fig: wiki plot} shows three vertical red lines corresponding to the model checkpoint after training for 3 epochs, 12 epochs, and 37 epochs. For simplicity of elaboration, we call them \textit{checkpoint early}, \textit{checkpoint mid}, and \textit{checkpoint best}. In terms of the \textit{checkpoint early}, the model is still under-fitted after only seeing the dataset a few times. Some high-level knowledge for general-domain text simplification has been learnt, but not enough to perform well. With respect to the \textit{checkpoint mid}, the model has established a firm understanding of the text simplification task in the general domain. Also, it is at the ``elbow point'' for the validation loss curve, meaning that the learning speed decreased significantly after this point. In terms of the \textit{checkpoint best}, the model has overfitted the training set, but its performance on the validation set is the best. We also incorporated the optimiser into the checkpoint for a smoother optimising process.

\begin{figure}[h!]
    \centering
    \includegraphics[scale=0.25]{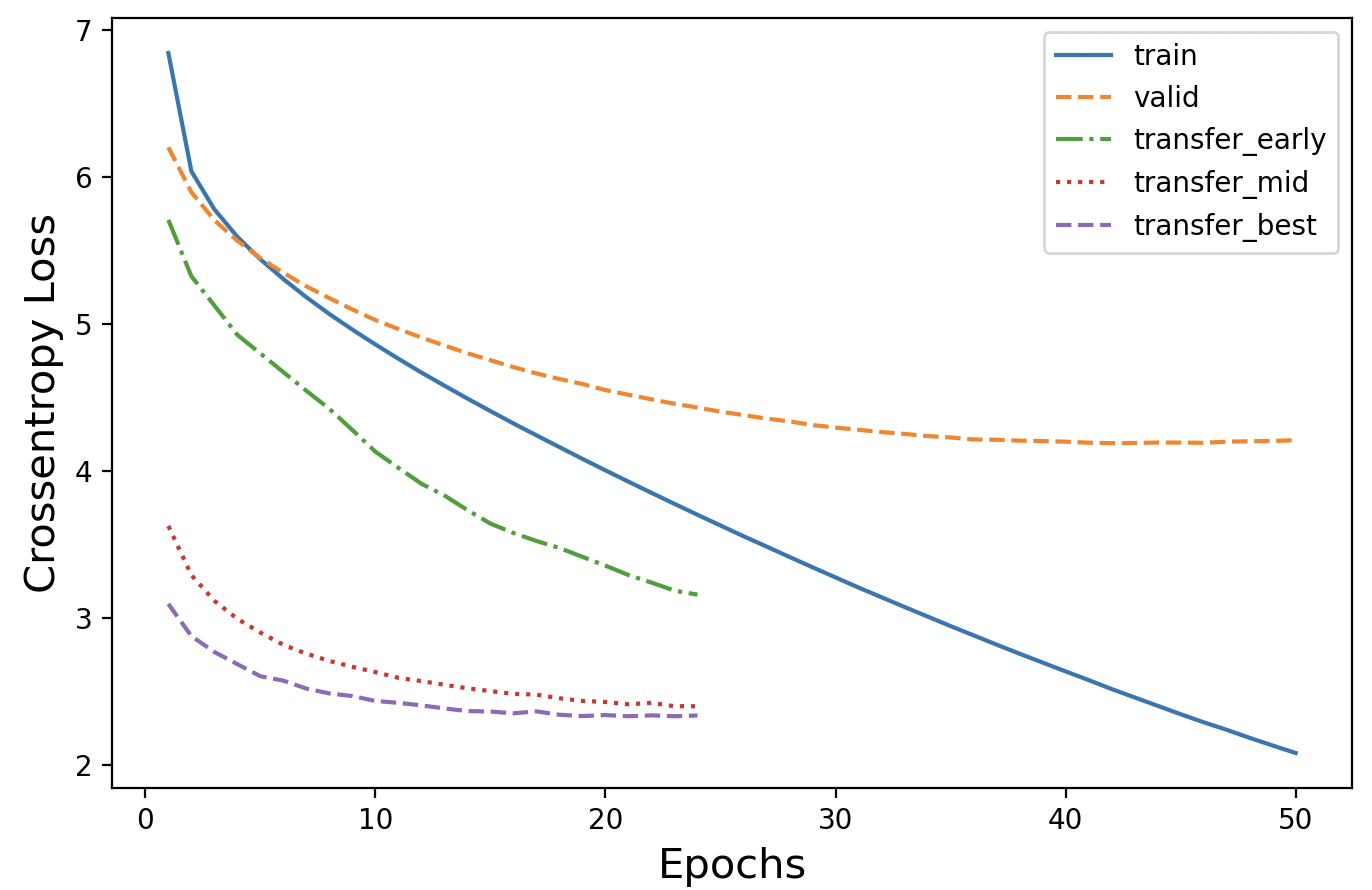}
    \caption{Transfer Learning Loss Curves}
    \label{fig:transfer curves}
\end{figure}

We adopted different transfer learning paradigms to explore the effect of transfer learning in the software documentation simplification task. Specifically, we started from the \textit{checkpoint early}, mid and best, and used these pre-trained checkpoints to train our models on the sw-data.  Figure~\ref{fig:transfer curves} contains the validation loss curves for all three different transfer learning paradigms and their comparison to the performance of the baseline sw.

 As seen in the figure, the cross-entropy loss curves for the three transfer learning strategies all have lower starting points. Moreover, the loss drops faster than the model trained solely on the sw-data. 24 epochs were trained on the three models and their loss in the validation set has reached the minimum. In terms of cross-entropy loss for the three transfer learning models, the \textit{checkpoint best} is the lowest, while the \textit{checkpoint early} is the highest.

However, loss in the validation set is merely an indicator of model performance. This metric suggests the uncertainty level of the model when decoding encoded sentences into their simplifications. The lower this metric, the more confident the model will be. However, as our sw-data dataset contains different writing styles and mask tokens, including URLs and code blocks, the model can find it difficult to generate more fluent sentences. Therefore, better performance in terms of the cross-entropy loss could happen just because the model learnt how to generate more fluent sentences from the checkpoint of the wiki-data. In this case, we need to look at the model performance in more detail.
\subsection{BLEU score Evaluation}
Sequence-to-sequence models have an exposure bias problem~\cite{ranzato2015sequence}. Therefore, we use the beam search method, which keeps track of the top k most probable candidate words during the data generation part. We choose the beam size k to be 5. 24 epochs were trained for the three transfer learning models. We also took snapshots of the models after every four epochs of training. For example, for the \textit{checkpoint early} model, we took the checkpoints after it was trained on sw-data for epochs of 4, 8, 12, 16, 20 and 24. We generate simplification text on two baselines, plus all the transfer learning model checkpoints on the test set. The generated texts will be investigated more thoroughly in later sections.

\begin{figure}[h!]
    \centering
    \includegraphics[scale=0.25]{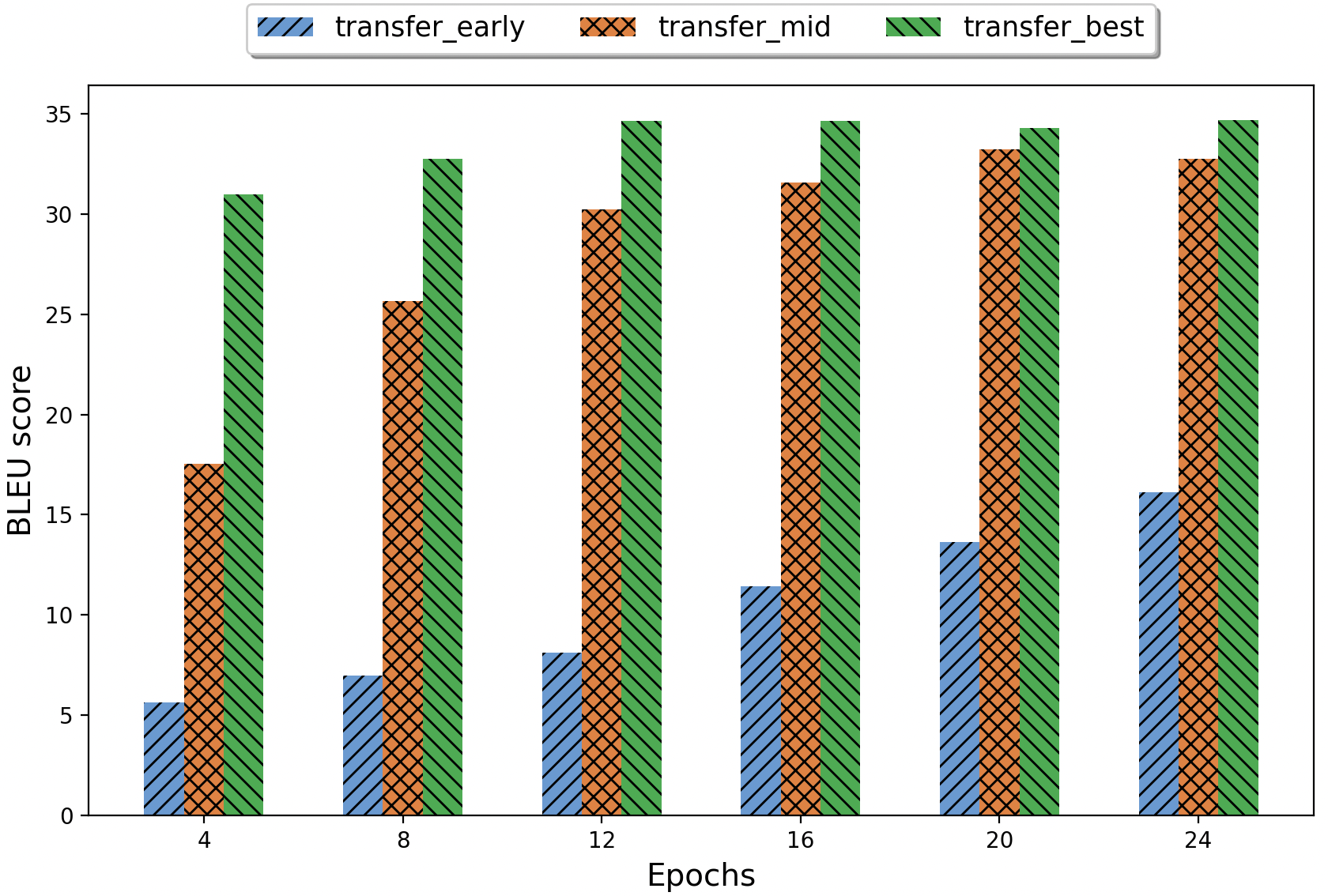}
    \caption{BLEU scores for Transfer Learning Models}
    \label{fig:BLEU transfer}
\end{figure}

We used the BLEU score to evaluate the quality of the texts generated in the last section. BLEU score measures the similarity of the generated text with its references and is widely used in machine translation tasks. It calculates the n-gram overlaps between the generated text and references. In our experiment, an equal weight of one quarter is given to unigram, bigram, trigram, and 4-gram to compute the BLEU score.
The BLEU scores for the baseline sw and the baseline wiki are 13.35 and 14.93. Figure~\ref{fig:BLEU transfer} shows the BLEU scores for all other transfer learning models. We also included an off-the-shelf simplification model from Nisioi et al~\cite{nisioi2017exploring}, which reached a BLEU score of 25.70.

\vspace{-1.5em}
As seen from the figure, the BLEU score for transfer\_early model starts at the lowest, and it increases gradually to 16.06 after training for 24 epochs, surpassing the BLEU score for both baseline wiki and baseline sw. For transfer\_mid model, the BLEU score increases significantly as the training on sw-data proceeds. It eventually reaches a BLEU score of 32.73.  On the other hand, the transfer\_best performance starts with the highest BLEU score and barely increases as the training proceeds. However, it still has the highest BLEU score of 34.68 in the test set. The BLEU score performance for the transfer\_mid model is slightly lower than that of the transfer\_best, except for the lower initial points. Furthermore, the BLEU scores for all three models are better than those for the two baselines.

\begin{tcolorbox}[left=1pt, top=1pt, right=1pt, bottom=1pt] \textbf{Summary of transfer learning:}
The baseline model directly trained on sw-data suffered from overfitting. By continuing training the checkpoints for wiki-data, models achieve a lower cross-entropy loss. Meanwhile, the transfer learning models surpass our two baseline models in BLEU score performance. The baseline wiki and baseline sw achieved BLEU scores of 14.93 and 13.35, while our best model achieves a BLEU score of 34.68.
\end{tcolorbox}



\section{Human Annotation}\label{human anno}
To further verify the effectiveness of the transfer learning approaches, We elaborate on the process of evaluation below.

\subsection{Procedure}

We selected four models to be evaluated in this phase, including three baseline models and one transfer learning model. The baseline models are the wiki-data trained model, sw-data trained model and the model from Nisioi et al.~\cite{nisioi2017exploring}, and we refer to as ``Baseline 1'', ``Baseline 2'' and ``Baseline 3''. The transfer learning model is further trained on the transfer\_best checkpoint using sw-data for another 24 epochs, which we refer to as ``Transfer''.

We adopted a similar method to that of Liu et al.~\cite{liu2019automatic} by first randomly selecting 100 original sentences from the sw-data test set, and generating texts using the four models. For each original sentence, there will be four versions of the simplification. For each of the 100 groups, we randomly shuffled the order of these versions to prevent annotators from discovering the patterns and making biased judgements. Also, if models generate the same new texts, they will be reduced to one piece of text for annotators to mark. This approach is used to avoid accidentally giving different marks to the same output.

We conducted a Prolific survey by dividing the 100 questions into 10 different surveys. Each survey was taken by three different participants. We performed a sequential survey release strategy. Specifically, for subsequent survey publications, individuals who had previously participated were excluded from the sample. In this case, participants cannot participate in the study multiple times. Meanwhile, we followed a study design similar to Nadi and Treude~\cite{nadi2020essential} by inserting a ``quality gate'' in a random position in the survey. We used the sentence ``The purple monkey dishwasher sang shenanigans on the moon with unicorns and marshmallow socks.'' consistently in all surveys as the ``quality gate''. This sentence has semantics clearly irrelevant to the reference sentence. We filtered out participants who did not give a semantic score below three for this sentence.

We used a similar evaluation metric as in related work~\cite{koptient2021fine}. Annotators are provided with the source sentences and their different simplified versions. During their annotation, they were asked to assess each generated sentence based on three evaluation criteria:

\begin{itemize}
    \item Simplicity: if the generated sentence is simpler than the original sentence.
    \item Semantic Similarity: if the generated sentence retains all semantics of the original sentence.
    \item Grammar: if the generated sentence is grammatically correct.
\end{itemize}

Likert scale is used to mark each of the aspects, with a score of 1 for strongly disagreeing and a score of 5 for strongly agreeing.


\subsection{Demographics of Annotators}
We first applied predefined filters from Prolific, which requires the participants to be in the employment sector of Information Technology (IT), while also being proficient in English. We asked the same question about employment again at the beginning of our survey and filtered out 45 participants who did not answer this question consistent with what they registered on Prolific. Our ``quality gate'' question further filtered out two participants.

In the survey, we asked them about their job roles and how many years they have worked in the IT field. For the 30 participants who passed all the filters and submitted their responses, they have on average 4.4 years of experience working in IT, with a maximum of twelve years and a minimum of half a year. Among the 30 participants, there are ten developers (33.3\%), seven IT support staff (23.3\%), four project managers (13.3\%), four data analysts (13.3\%), four administrators (13.3\%) and one search quality rater (3.3\%).


We computed their level of agreement on the annotation results, which reached 0.42 for Krippendorff's alpha~\cite{krippendorff2004reliability} coefficient overall, with semantics, grammar and simplicity at 0.53, 0.37 and 0.32, respectively.  This indicates that annotators reached some agreement, while some of the results are subjective at the same time.

\vspace{-1em}
\subsection{Results}
 A good simplification of a sentence should not only be ``simpler'' than the original sentence, but should also preserve semantics and be grammatically correct. We provide one counter-example, where the original sentence is ``Note: to create a debug build of the building files, pass the $--debug$(or $-d$) switch when running the either configure or build command'' and the simplified sentence by Baseline 1 is ``To create a compact build of the unlimited file, pass the award tells the story of the game''. In this example, two annotators gave a semantic score of one, but a simplicity score of four and five, respectively. However, because of the subjectivity of the annotation process and the diversity of the participants, the last annotator gave a score of two in simplicity for this sentence. This sentence does not preserve any meaning from the original sentence. We argue that these generated sentences that either fail to preserve the meaning or are grammatically wrong are not usable. Therefore, we define a ``good'' sentence as one with a semantic score and a grammar score of at least four.

\begin{figure}[h!] 
\begin{subfigure}{0.21\textwidth}
\includegraphics[width=\linewidth]{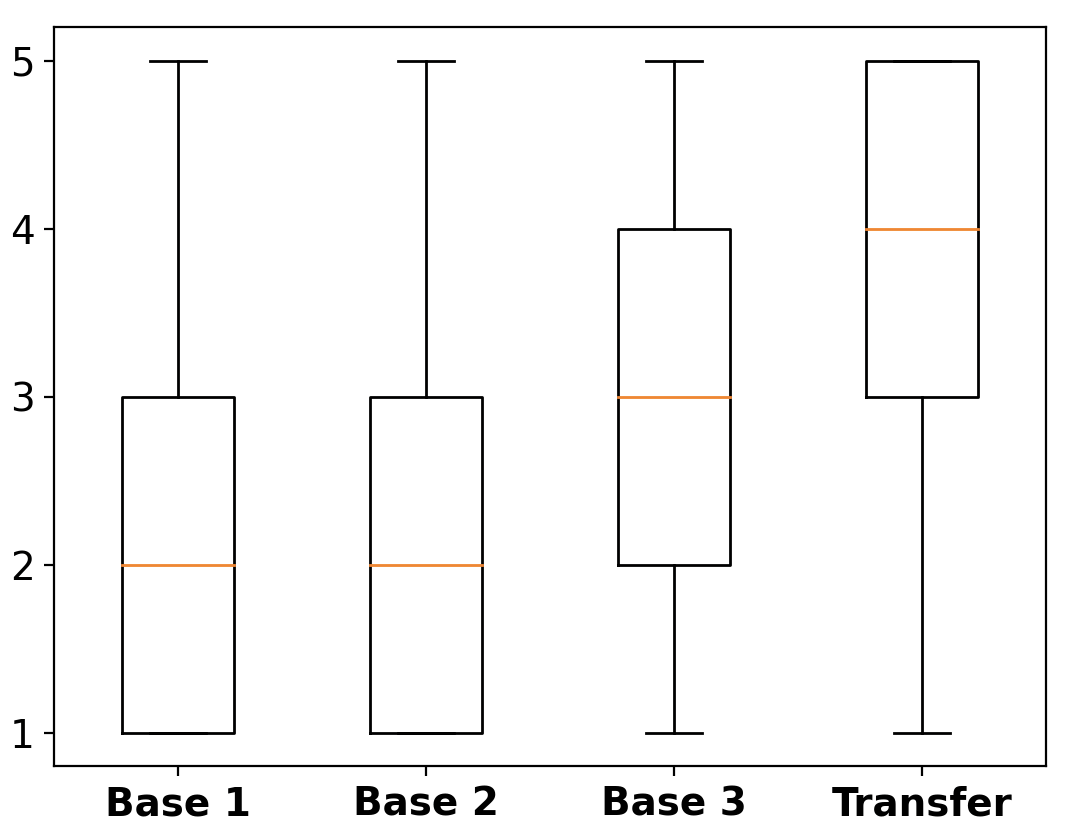}
\caption{Semantic box-plot} \label{fig:first}
\end{subfigure}\hspace*{\fill}
\begin{subfigure}{0.21\textwidth}
\includegraphics[width=\linewidth]{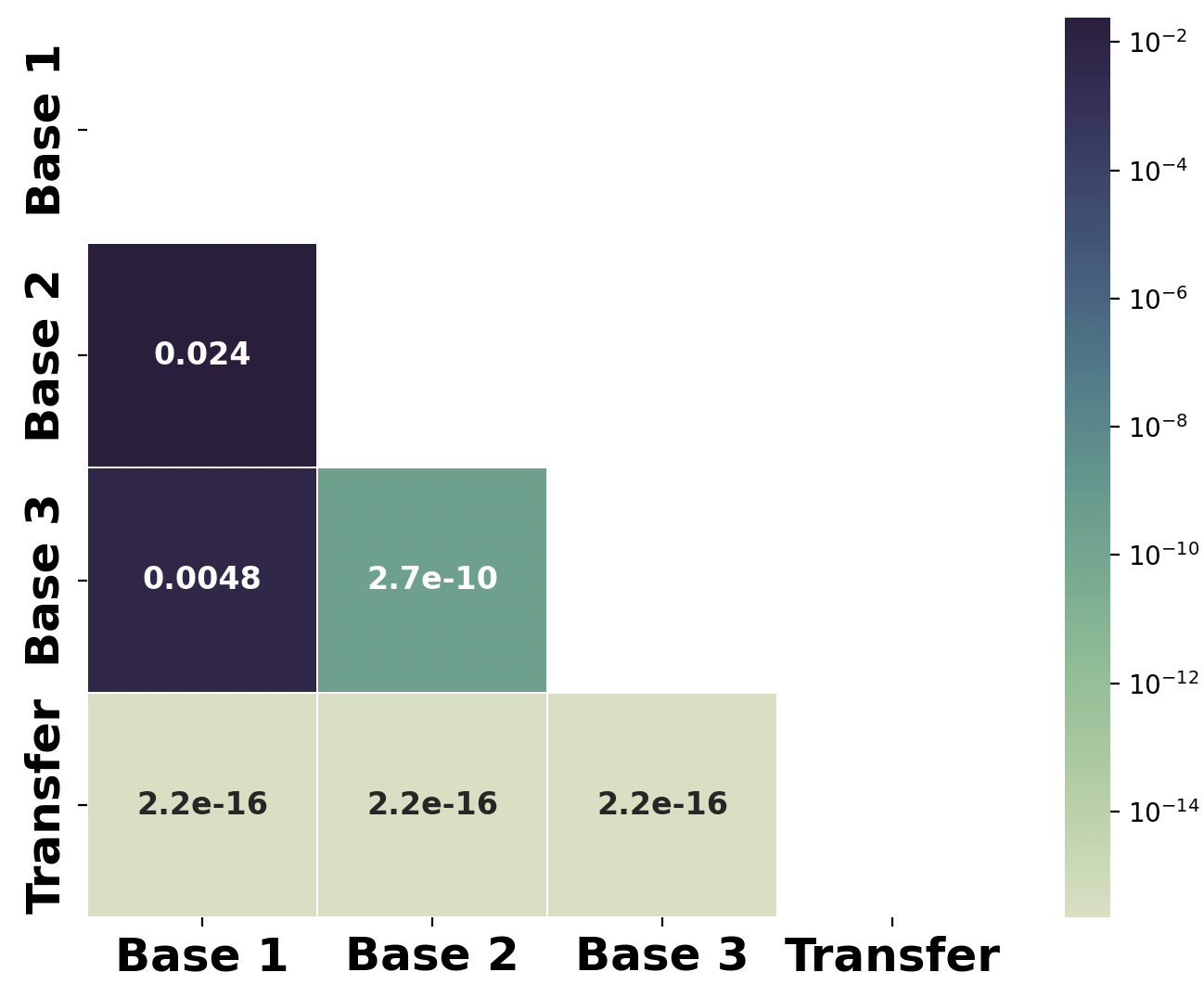}
\caption{Semantic p-values} \label{fig:heat_first}
\end{subfigure}

\medskip
\begin{subfigure}{0.21\textwidth}
\includegraphics[width=\linewidth]{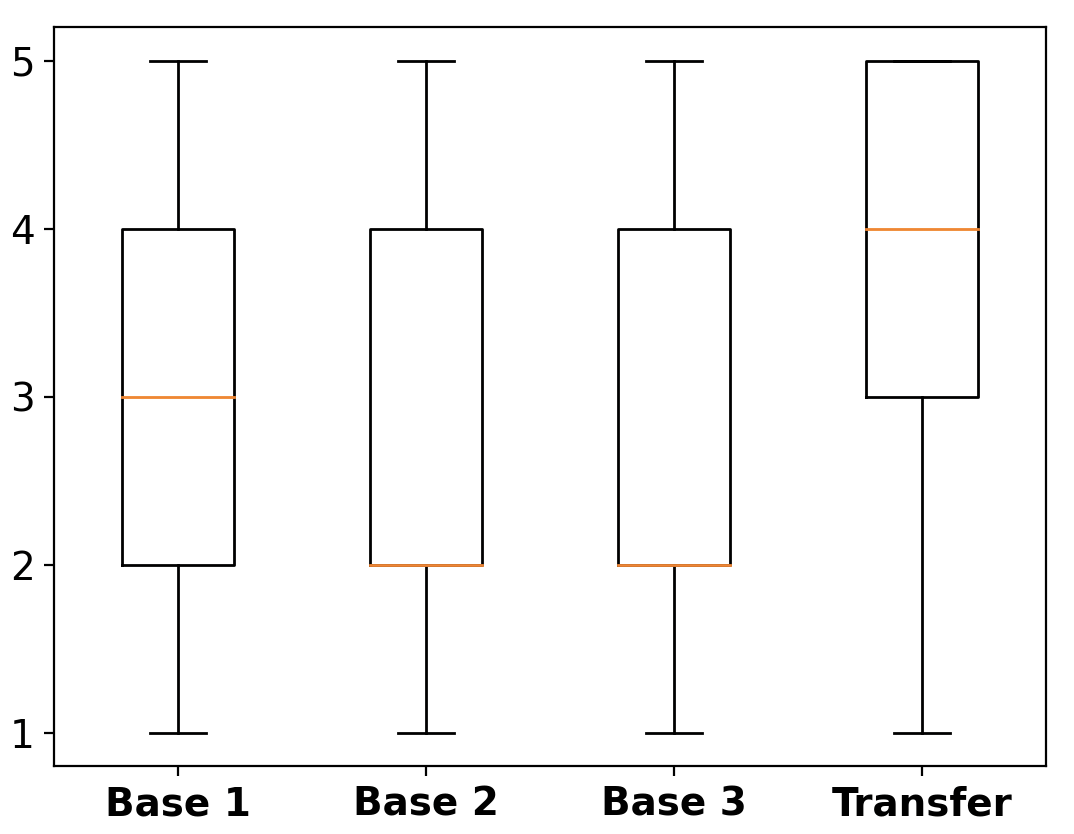}
\caption{Grammar box-plot} \label{fig:heat_second}
\end{subfigure}\hspace*{\fill}
\begin{subfigure}{0.21\textwidth}
\includegraphics[width=\linewidth]{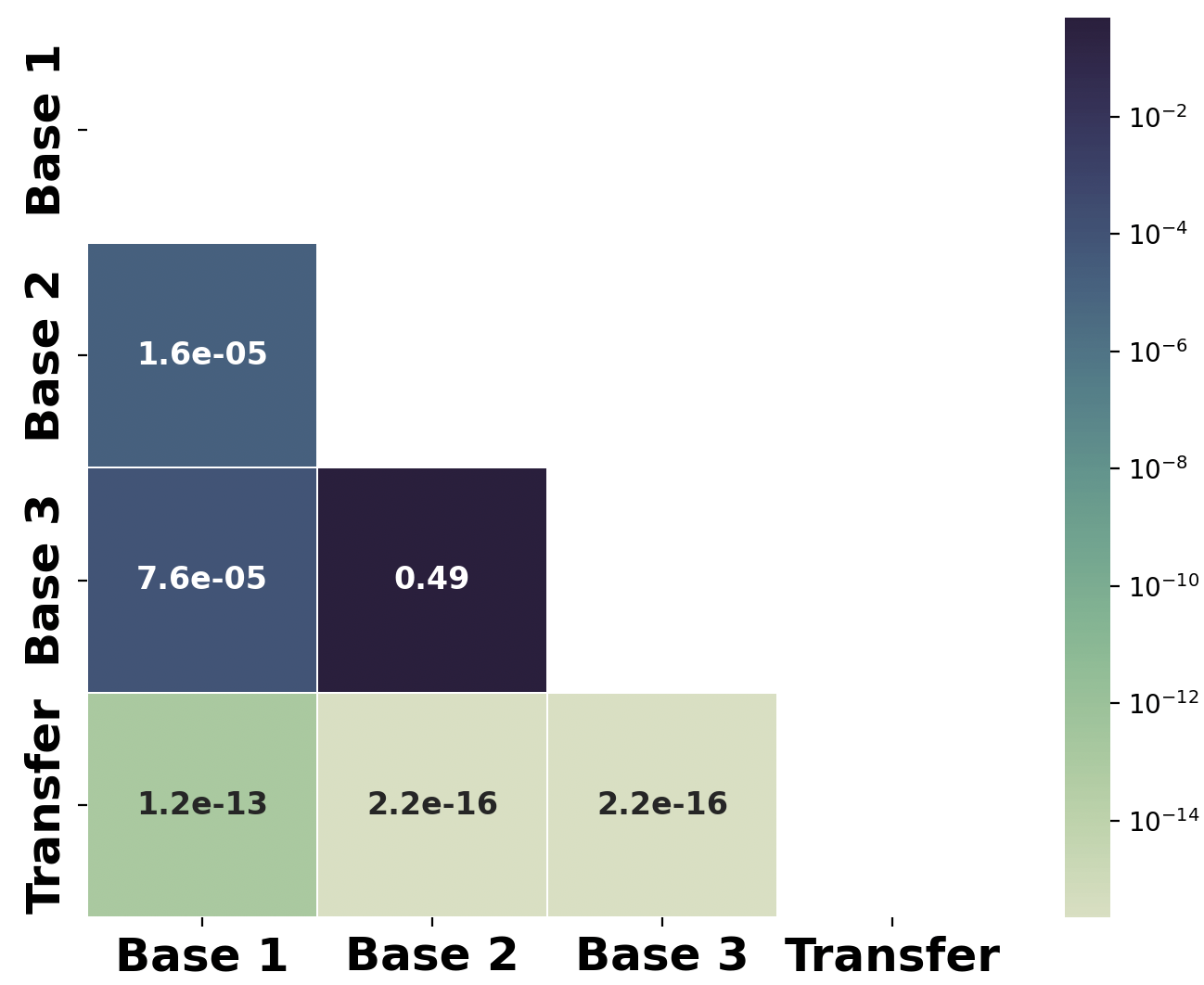}
\caption{Grammar p-values} \label{fig:heat_third}
\end{subfigure}

\medskip
\begin{subfigure}{0.21\textwidth}
\includegraphics[width=\linewidth]{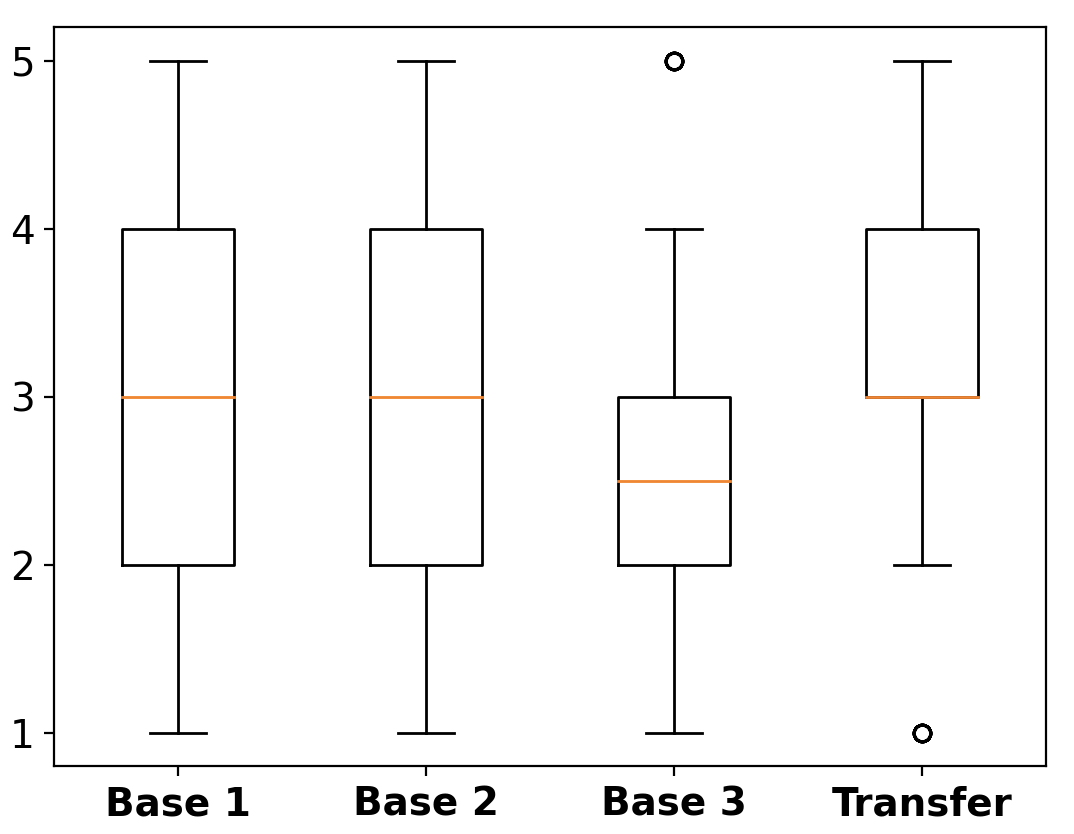}
\caption{Simplicity box-plot} \label{fig:third}
\end{subfigure}\hspace*{\fill}
\begin{subfigure}{0.21\textwidth}
\includegraphics[width=\linewidth]{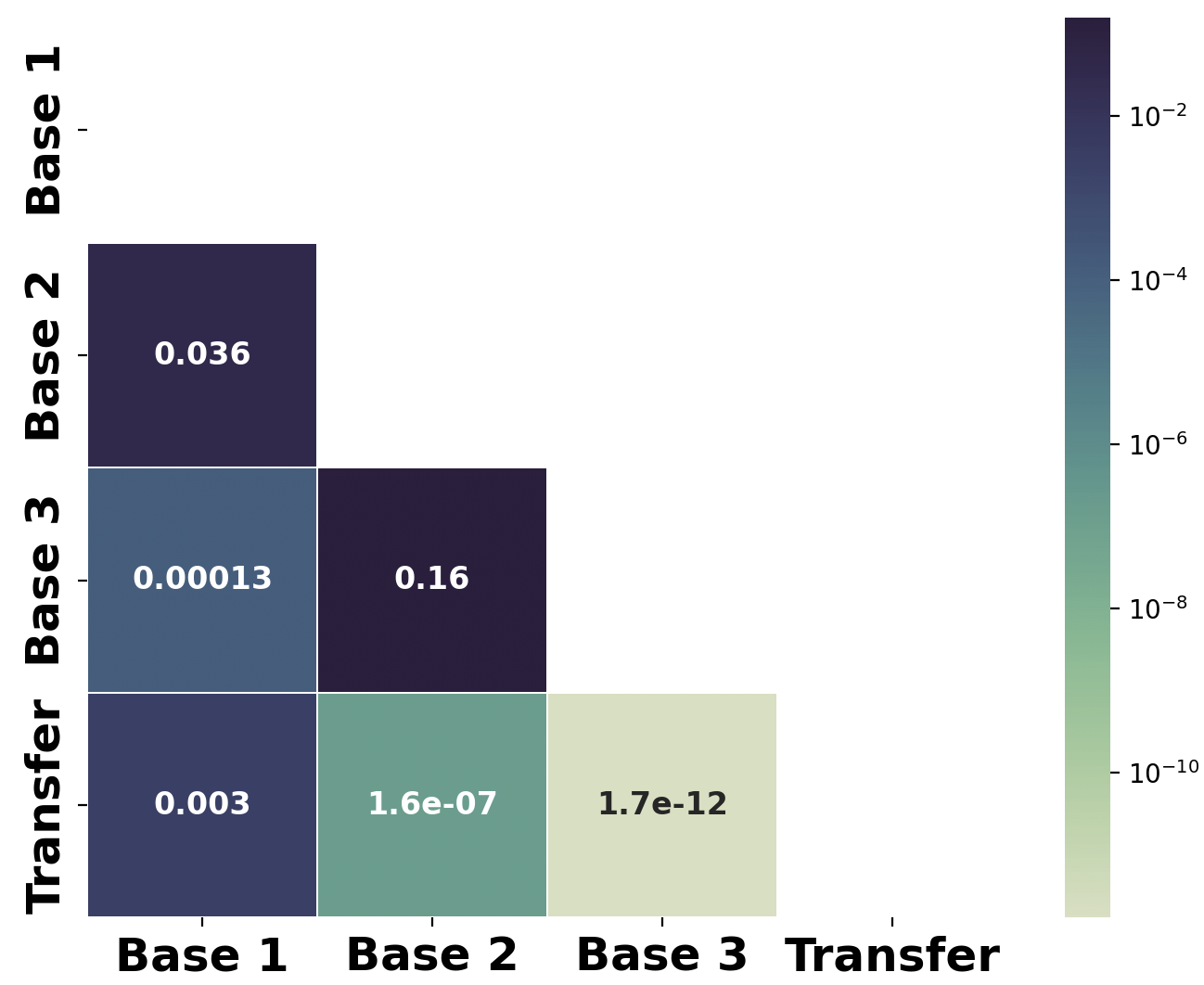}
\caption{Simplicity p-values} \label{fig:f}
\end{subfigure}

\caption{Score Box-plot and Pairwise Wilcoxon p-values} \label{fig:survey stats}
\end{figure}

Table~\ref{tab:likert scores} shows the average Likert scores for all models in the three metrics mentioned above, as well as the number of instances with semantic score, grammar score or both no less than four. The left column in Figure~\ref{fig:survey stats} shows the box plot for the distribution of these three aspects. We brief the Baseline 1 to three with the name ``Base'' 1 to 3 to clearly present them in the figure.

\begin{table*}[h!]
    \centering
        \caption{Likert score for Models}
    \begin{tabular}{llrrrrrr}
       \toprule
    & & Semantics & Grammar & Simplicity & \#Semantics$\geq$4 & \#Grammar$\geq$4 & \#Good  \\
     \midrule
     1 & Baseline 1 & 2.427 & 3.097 & 3.017 & 71 & 136 & 54 (18.0\%)\\
     2 & Baseline 2 & 2.113 & 2.637 & 2.790 & 45 & 90 & 33 (11.0\%) \\
     3 & Baseline 3 & 2.730 & 2.700 & 2.557 & 108 & 98 & 64 (21.3\%)\\
     4 & Transfer &\textbf{3.907} & \textbf{3.873} & \textbf{3.350} &\textbf{219} & \textbf{211} & \textbf{182 (60.7\%)}\\
     \bottomrule
    \end{tabular}
    \label{tab:likert scores}
\end{table*}

\begin{table*}[h!]
    \centering
    \caption{Generation Examples}
    \begin{tabular}{lp{7.5cm}p{7.5cm}}
    \toprule
    & Original & Simplified\\
    \midrule
    Example 1 &
  If you're interested in using speaker notes, reveal.js comes with a node server that allows you to deliver your presentation in one browser while viewing speaker notes in another.
     & 
    reveal.js comes with a speaker notes plugin which can be used to present per-slide notes in a separate browser. \textbf{(by Transfer model) } \\
    \midrule
    Example 2  & 
     gldispatch/ contains code for libgldispatch, which is responsible for dispatching opengl functions to the correct vendor library.&
     gldispatch/ contains code for libgldispatch. \textbf{(by Transfer model)}
    \\
    \midrule
    Example 3 & When a collection is typed as Seq[String], so might have linear access like List, but actually is a WrappedArray[String] that can be efficiently parallelized, but can be efficient with scala parallel collections. & When a collection is typed as Seq[String], so might have linear access like List, but actually is a WrappedArray[String] that can be efficient.(\textbf{by Transfer model}) \\
    
    \bottomrule
    \end{tabular}
    \label{tab:generation example}
\end{table*}

Overall speaking, Baseline 2 performs the worst in the semantic and grammar aspects, and second to last for simplicity aspect. Although Baseline 1 is able to generate grammarly moderately acceptable sentence, because it is not trained on sw-data, it fails to generate the domain-specific text, causing the loss of semantic. The simplicity score for Baseline 1 is regarded at an acceptable scale. In terms of Baseline 3, it performs not ideal on all three aspects.

For Transfer model, its scores largely surpass the rest of the models. It is capable of generating more grammatically correct sentences that preserve semantic meanings. Also, it has an average simplicity score of 3.35, indicating that the generated sentences are on average simpler than the original ones.

We can have a more detailed comparison via the box plots in Figure~\ref{fig:survey stats}. Baseline 3 performs better in terms of semantic preserving, while Transfer model consistently performs better in all three aspects. In terms of the simplicity aspect, Baseline 1 and Baseline 2 occasionally generate satisfying simplification that is above a score of three, while Baseline 3 consistently performs the worst.


{From a different perspective, the Transfer model can generate 219 sentences that well preserve the original sentence semantics and 211 sentences that are grammarly correct out of the total 300 sentences according to the survey result. Overall, more than 60\% of the sentence generated by the Transfer model is ``good'', while most of the sentences generated by other models are barely usable.

Additionally, we performed statistical tests to verify whether the differences shown above are statistically significant. 
We performed Wilcoxon signed rank tests~\cite{rey2011wilcoxon} on all pairs of available data on semantic, grammar, and simplicity scores. The right column in Figure~\ref{fig:survey stats} includes the heatmaps for all pairs of models, with the calculated p-values listed in cells. One important note is that pairs that are identical to the original sentence (22 out of 100) were not considered when performing the statistical test for simplicity score.

 

If taking a significance level of 0.05, we can see that the score for Transfer model is statistically better than all other models in every aspect. Meanwhile, the differences in grammar score and simplicity score between Baseline 1 and Baseline 2 are statistically significant. This indicates that by applying transfer learning on the wiki-data checkpoint, the model could learn how to generate more fluent sentences that are grammarly correct and perform meaningful simplification.

To further explain these results, we analysed interesting cases in more detail. We found that baselines 1 and 2 have an interesting distribution in terms of simplicity. Although they have relatively lower scores of 2.790 and 3.017, around 25\% sentences are higher than four for both Baseline1 and Baseline 2. We further divided the annotation source sentences into one group containing at least two masked special components and another with less than two. For Baseline 1, this results in an average simplicity score of 3.18 for sentences with less than two masked tokens, and the average score drops to 2.74 when there are at least two masked tokens. Meanwhile, the performance for Baseline 2 increases from 2.63 to 3.04 for these two groups.

The decrease in performance for the model trained on wiki-data is reasonable as there are no masked components in general-topic text, and the domain transition would introduce a performance drop. In contrast, the model trained on sw-data sees an increase in performance when more masked special tokens appear in the sentences. Meanwhile, Baseline 1 performs better in generating grammatically correct sentences. Therefore, we argue that wiki-data checkpoint brings more coherent sentences, while further training on sw-data enhances the preservation of semantics and gives the model better versatility with the masked special tokens.

Lastly, we provide three examples  of the simplification generated by the Transfer model in Table~\ref{tab:generation example}. The first example omits unimportant details while making the sentence easier to understand. The second example discards the second half of the sentence. The third example omits a technical part of ``WrappedArray'' by not mentioning its parallelization. By rewriting and omitting parts of the sentences in a way that does not severely interfere with the semantics, the Transfer model can provide sentences perceived as simpler by our annotators. The simplicity scores for these three examples were 4.67, 4 and 3.67, on average, while their semantic scores ranged between 4 to 5, 3 to 4 and 2 to 4, respectively. 

\vspace{-0.5em}

\begin{tcolorbox}[left=1pt, top=1pt, right=1pt, bottom=1pt] \textbf{Summary of human annotation results:}
Our best model (Transfer) consistently outperforms three baselines in all three aspects.
Wiki-data checkpoint enhances the coherence and grammar of the generated sentences, while further training on sw-data improves the preservation of semantics and gives the model better versatility with the masked special tokens.
\end{tcolorbox}
\vspace{-1em}

\subsection{Analysis on Identical Sentences}

As the sw-data simplification is less prominent than wiki-data, models sometimes learn to predict sentences identical to the original ones. In the 100 sentences used for our annotation, we found that transfer learning models with higher scores tend to generate more replications. Specifically, for the 100 cases, Baseline 2 did not generate any replication, while Baseline 1 and Baseline 3 generated 2 replication each. On the other hand, Transfer generated 22 replications. 
For example, for the original text ``The chain method takes one argument: m.chain(f), f must be a function which returns a value if f is not a function, the behaviour of chain is unspecified.'', Transfer model generated identical output, while Baseline 1 generated a shorter sentence with ``The variable method takes one argument is a function which returns a value if a mathematical is not a function''. The Baseline 1 generated sentence omits many details and barely retains semantic information. This level of simplification is not practical for developers as details are missing and semantics are degraded. However, it is hard for the model to learn to simplify effectively in each scenario, especially when the sw-data contains more replications. This motivates us to incorporate domain-specific rules in our future work.

\vspace{-0.6em}
\section{Treats to Validity}
We consider threats to the validity of our study in this section.

The first threat is that we mined software repositories from the first GitHub index and did not collect repositories created after 2017. We believe that more README files in older repositories need to be simplified, as different techniques were used back then, and more old repositories have gone through simplification updates compared to recent ones. Collecting datasets from different creation periods of time could potentially give different simplification results. 

Second, additional context-related components, such as package requirements, could potentially be masked when assigning new tokens. However, these components are usually embedded in plain text and it would require sophisticated regular expression tools for extracting them. Regular expressions are known to be noisy when processing plain text, and this is not our main purpose in this paper. Therefore, a better regular expression tool to preprocess the text could yield different results on the simplification dataset.

Third, due to the lack of computing resources, we did not extensively tune the hyperparameters on the models, which could lead to overfitting and suboptimal solutions. However, given that the performance gap in the same set of hyperparameters among models is quite obvious, the overall design would not be compromised.

Lastly, we acknowledge that the BLEU score is not an ideal indicator for simplification tasks, and this motivates us to perform a survey with human participants. Although our annotation results revealed that users find that the transfer learning model generates the most satisfying result, our study does not provide evidence on the impact of simplification on comprehension tasks.

\section{Implications and Future Work}

The simplification of README files has significant implications from several perspectives. First, from a newcomer's perspective to a repository, a simpler version of the README files has the potential to help newcomers understand the project structure faster and mitigate the technical barriers. Second, from the perspective of repository owners, an easy-to-read README file could enhance the repository's potential to attract more users and participants. Third, from the perspective of document writers, the recommended simplified version of their text could help them to take care of certain groups of readers when composing the draft.

In addition, as README files usually take the role of project walkthrough and tutorial, similar ideas of simplification could be applied to software teaching materials and other relevant versions of tutorials. This work explores the simplification operations from README documentation and fills the gap between general-style text simplification and domain-specific simplification. Our transfer learning approach provides a direction for using general-style simplification knowledge to compensate for the lack of knowledge in domain-specific simplification settings. 

Although we performed a Prolific survey of people with IT backgrounds, one limitation of is that we did not collect longitudinal evidence on the effects of simplified documentation on different stakeholders in open source. Different people interpret ``simple'' differently, which is indicated by the moderate Krippendorff's alpha score from the Prolific survey. In addition, documents from different programming languages may need different simplifications because of the techniques they use and the communities they are in.

 Moreover, Wikipedia data is found to be biased in culture, gender and other perspectives~\cite{callahan2011cultural, wagner2015s}. Although we did not investigate this issue, using the transfer learning model trained on this data, biased use of words might be carried forward. This could be detrimental, especially when some communities are found to be more toxic in language~\cite{raman2020stress}. Therefore, more investigation into this issue is an important direction for future work.

In terms of future work, people with different levels of expertise may find different levels of detail easier. For example, entry-level developers might find a comprehensive document easier to understand. At the same time, people with more expertise might need just enough documents that are ``to the point''. This situation also applies to different job roles and ages. Therefore, we intend to perform more user-centred studies to elucidate how different groups perceive the concept of ``simplification''.

In addition, more empirical studies on how README files are updated for readability and simplicity purposes are also in the future direction, specifically: (1) What repositories tend to include more simplification operations? (2) What aspects are the simplification operations focusing on? (3) What triggers the simplification operation? Through qualitative and quantitative studies, we could summarise the gap between people's perceptions and common practices for READMEs simplification, providing more concrete advice on the aspects to pay attention to during documentation writing.

Lastly, to automate the process of human-centred software document simplification, higher-quality data, different metrics, and simplification rules could all be incorporated into the system. Also, in the era of Large Language Models (LLM)~\cite{brown2020language}, we could consider prompt engineering LLM~\cite{zhou2022large} for performing this task.

\section{Conclusion}
In this paper, we collected README files from GitHub and used the BERT sentence alignment algorithm and multiple heuristic filters to construct a README files simplification dataset. Then we trained a transformer model on both wiki-data and sw-data and performed transfer learning by continuing training the model on sw-data from the wiki-data trained checkpoints. After that, we performed a Prolific survey, asking people with IT background to annotate 100 groups of sentences generated by different models from the perspective of semantic preserving, grammarly correctness and simplicity. The best transfer learning mode outperforms the baselines in both the automatic evaluation of BLEU score and the human evaluation. We found that the transfer learning model learns to perform meaningful simplification behaviours to the sentences while preserve the original meaning of the sentences.

\section{Data Availability}
The replication package is at \url{https://zenodo.org/record/8265001}.

\printbibliography
\end{document}